# Effects of language mismatch in automatic forensic voice comparison using deep learning embeddings

Dávid Sztahó[1], Attila Fejes[2]

[1]Budapest University of Technology and Economics, 1117 Budapest, Magyar tudósok körútja 2., Hungary

[2]University of Public Service Doctoral School of Law Enforcement, Ludovika tér 2., 1083 Budapest, Hungary

szaho.david@vik.bme.hu, fejes.attila@nbsz.gov.hu

## Abstract

In forensic voice comparison the speaker embedding has become widely popular in the last 10 years. Most of the pretrained speaker embeddings are trained on English corpora, because it is easily accessible. Thus, language dependency can be an important factor in automatic forensic voice comparison, especially when the target language is linguistically very different. There are numerous commercial systems available, but their models are mainly trained on a different language (mostly English) than the target language. In the case of a low-resource language, developing a corpus for forensic purposes containing enough speakers to train deep learning models is costly. This study aims to investigate whether a model pre-trained on English corpus can be used on a target low-resource language (here, Hungarian), different from the model is trained on. Also, often multiple samples are not available from the offender (unknown speaker). Therefore, samples are compared pairwise with and without speaker enrollment for suspect (known) speakers. Two corpora are applied that were developed especially for forensic purposes, and a third that is meant for traditional speaker verification. Two deep learning based speaker embedding vector extraction methods are used: the x-vector and ECAPA-TDNN. Speaker verification was evaluated in the likelihood-ratio framework. A comparison is made between the language combinations (modeling, LR calibration, evaluation). The results were evaluated by $Cllr_{min}$ and EER metrics. It was found that the model pre-trained on a different language but on a corpus with a huge amount of speakers can be used on samples with language mismatch. The best EER with speaker enrollment was 1%. The effect of sample durations and speaking styles were also examined. It was found that the longer the duration of the sample in question the better the performance is. Also, if multiple recordings are present from the suspect, there is no real difference if various speaking styles are applied.

# Introduction

Speaker identification (SI) and verification (SV) (together: speaker recognition) have been studied for a long time and have a still growing literature (Hanifa et al., 2021). A large number of studies have been dealt with the field, resulting in many methods. State-of-the-art technologies vary across time periods as new and new ideas are examined. Until recently, the i-vectors were considered as the state-of-the-art technique for speaker recognition (Hansen and Hasan, 2015), but with the emergence of deep learning methods and the appearance of huge speech corpora novel classification and feature extraction methods are developed (such as d-vectors (Variani et al., 2014), j-vectors (Chen et al., 2015), x-vectors (Snyder et al., 2017) and ECAPA-TDNN networks (Desplanques et al., 2020)).

In speaker identification the task is to identify an unknown speaker from a set of already known speakers: find the speaker who sounds closest to the test sample. Closed-set (or in-set) scenario is when all speakers within a given set are known. On the other hand, open-set (or out-of-set) speaker identification is when the set of known speakers may not contain the potential test subject (Sztahó et al., 2021).

In speaker verification, we verify if a speaker is who he or she claims to be by comparing two speech samples/utterances and evaluating if the speakers of the two samples match (Sztahó et al., 2021). This is traditionally done - in general forensic voice comparison practice - by comparing the test sample to a sample or samples of the given speaker and a universal background model (Reynolds and Rose, 1995). Another way of comparing if a speaker pair is of same origin is to score the pairs as 'same' or 'different' and create a model accordingly. This is feasible for sample-wise comparisons, because a model can be trained on a dataset with sample pairs to predict whether the speakers are the same instead of using a UBM. It follows from this definition that technically, forensic speaker comparison belongs to the speaker verification scheme, although the 'known' speaker is not specifically known in this case. A voice sample may be associated with an assumed speaker. The goal is to verify if this speaker's identity (suspect) matches another unknown speaker's identity (offender). Often, the goal of the forensic voice comparison is also to verify if two unknown speaker's identity is the same. In practice, this verification is done with the same method.

There is a paradigm shift in forensic sciences and practice (Morrison, 2011a, 2009; Saks and Koehler, 2005) that makes the automatic and semi-automatic evaluation of evidence available using various modalities and kinds of measurements (such as DNA, fingerprint) (Bazen and Veldhuis, 2004; Matz and Nielsen, 2005). This new paradigm, called likelihood-ratio (LR) framework, supports a processing pipeline easily computable across multiple evidence types. Forensic voice comparison, where speaker recognition techniques are adapted to the requirements of the framework, is such a field, where this new paradigm can be adapted. It produces a probability ratio of same and different speakers for an evidence (Kelly et al., 2019a; Mandasari et al., 2011; Morrison, 2011b).

Considering a forensic voice comparison system, we can evaluate the same and different speaker origin comparison in two ways (Poddar et al., 2018): (1) multiple samples are available from unknown and known speakers and (2) we can compare the samples pairwise. It naturally follows that the first scheme can achieve higher accuracy. However, there are many cases when multiple samples are not available for comparison (only a recording excerpt is available). There have been multiple studies using short utterances (Li et al., 2016; Min Kye et al., 2021; Rohdin et al., 2020; Wang and Hansen, 2022), but they commonly lack the forensics nature: datasets not following a strict protocol (Morrison et al., 2012) or they used techniques already outperformed by deep learning ones in regular speaker recognition. This study aims to investigate this scenario in two ways: only one sample is available for the unknown speaker and (i) only one or (ii) multiple samples are available for the known speaker. An example for (i) is that if one speech sample is available for both the offender and the suspect, and for (ii) is if one sample is available for the offender but multiple samples can be recorded for the suspect. Therefore, in this study we focus on pairwise comparison of samples.

In the LR framework of forensic voice comparison, we have to investigate two hypotheses: (1) "What is the possibility that the sample in question originates from the suspect?" and (2) "What is the possibility that the sample in question originates from a randomly selected speaker of a background population?". The ratio of these expressions expresses the strength of the evidence (Eq. 1). LR is the likelihood-ratio, $E$ is the evidence, $H_{so}$ is the hypothesis of same-origin speakers, $H_{do}$ is the hypothesis of different-origin speakers.

Eq. 1. $$LR = \frac{P(E|H_{so})}{P(E|H_{do})}$$

There are multiple large-scale corpora available (Chanchaochai et al., 2018; Nagrani et al., 2020; Panayotov et al., 2015) for speaker recognition purposes and the NIST speaker recognition challenge (Sadjadi et al., 2022) is also arranged permanently. However, there are specific needs for a forensic voice comparison system (Morrison et al., 2012) that these corpora do not fulfill. The present study uses a speech dataset developed for forensics to evaluate the speaker verification systems and to analyze their sensitivity to sample lengths and speaking style. The goal here is to compare the performance of state-of-the-art deep learning feature extraction models pre-trained on the VoxCeleb dataset with models trained on a low-resource language and evaluate them on the low-resource language. This will show the usability of pre-trained models with language mismatch (between the trained model and test samples) for forensic voice comparison for institutes like public services for countries where sufficient speech data cannot be collected to appropriately train a deep learning model.

Feature extraction methods based on deep learning (like the TDNN architectures) have shown performance superior to former approaches (GMM-UBM, i-vector). However, these techniques need a large amount of training data to get appropriate models. Low-resource languages lack the available data to train such models. Thus, the possibility comes naturally to use pre-trained models (trained on insanely large English datasets available) for speaker recognition for smaller languages. Kleynhans and Bernard (Kleynhans and Barnard, 2005) found a language dependent tendency, but they use a technology already outdated for speaker verification. Nevertheless, the language dependency of the deep learning feature extraction methods (such as the one used in this study) can be low due to the huge amount of data they are trained on. There have been studies investigating this cross-language scheme, but none of them fits into the forensic voice comparison framework. Li and his colleagues (Li et al., 2017) uses synthesized speech for evaluation and the comparison is pairwise. In the study of Chojnacka and his colleagues (Chojnacka et al., 2021), multilingual experiments were done with a multilingual training dataset which is not the main practice we would like to investigate. The language dependency of the i-vector technology was already studied (Misra and Hansen, 2014; Vaheb et al., 2018), but newer deep learning techniques already outperformed the older i-vector. Fabien and Motlicek (Fabien and Motlicek, 2021) studied the performance of the x-vector models in forensic scenarios but with acted speech and the study was not focusing on multilingual effects (although the dataset was multilingual). The study of Skarnitzl and his colleagues (Skarnitzl et al., 2019) is forensics specific and evaluates a multilingual scenario, but uses an earlier version of VOCALISE (Kelly et al., 2019b) system that is based on the outdated i-vector. They found controversial results in cross-language evaluations.

There are certain factors in speech material that can affect the effectiveness of voice comparison. It may be important to know if speaking style and sample duration mismatch deteriorate the performance. Few studies have been aimed to see if these factors really matter (Afshan et al., 2020; Afshan and Alwan, 2022; González Hautamäki et al., 2019), although if they do, it can add bias to the evaluation of the evidence. In the present study we use three speaking styles available in the dataset developed for forensic needs and compare the results depending on sample durations.

In this study we examine the followings: (1) how a pre-trained deep learning speaker embedding model performs on a low-resource language other than the one the model is trained on; (2) is there a performance increase and how large it is if multiple samples are available from the known speaker (suspect); (3) how do the performance metrics depend on the length of the samples and (4) how do the performance metrics depend on the speaking style (available in the dataset). The results can help forensic science services or institutes services planning forensic voice comparisons.

The paper is structured as follows: in the next Section we give a description on the methods, evaluation metrics and schemes used throughout the study. It is followed by the presentation of the results. In the next Section, we give a discussion about the obtained evaluations with a short concluding thoughts.

# Embedding models

Two techniques were used in this study to extract embedding vectors for speech samples as features: the x-vector and ECAPA-TDNN. These methods get a voice sample as input and give a vector as output that can be imagined as a vector representation of the speaker in the sample. Deep learning based feature extraction models were applied using the SpeechBrain toolkit (Ravanelli et al., 2021). In this study, three embedding models are used to evaluate cross-language speaker verification schemes. In order to compare the models pre-trained on the VoxCeleb dataset, the parameters of the custom-trained models followed the recipe of these pre-trained models. The pre-trained models were downloaded from Huggingface[1,2].

## The x-vector

The deep learning based feature extraction method called x-vector was developed primarily for speaker verification (Snyder et al., 2017). It is based on a multiple layered DNN architecture (with fully connected layers) with different temporal context at each layer (which they call 'frames'). Due to the wider temporal context, the architecture is called time-delay NN (TDNN). The TDNN embedding architecture can be seen in Figure 1 and Table 1.

*Table 1. The x-vector DNN layer architecture* (Snyder et al., 2017). *It contains the layers, contexts and the input, output dimensions.*

| Layer | Layer Context | Total Context | Input x Output |
|---|---|---|---|
| frame1 | [t − 2, t + 2] | 5 | 120 × 512 |
| frame2 | {t − 2, t, t + 2} | 9 | 1536 × 512 |
| frame3 | {t − 3, t, t + 3} | 15 | 1536 × 512 |
| frame4 | {t} | 15 | 512 × 512 |
| frame5 | {t} | 15 | 512 × 1500 |
| stats pooling | [0, T} | T | 1500T × 3000 |
| segment6 | {0} | T | 3000 × 512 |
| segment7 | {0} | T | 512 × 512 |
| softmax | {0} | T | 512 × N |

---

[1] x-vector: https://huggingface.co/speechbrain/spkrec-xvect-voxceleb
[2] ECAPA-TDNN: https://huggingface.co/speechbrain/spkrec-ecapa-voxceleb

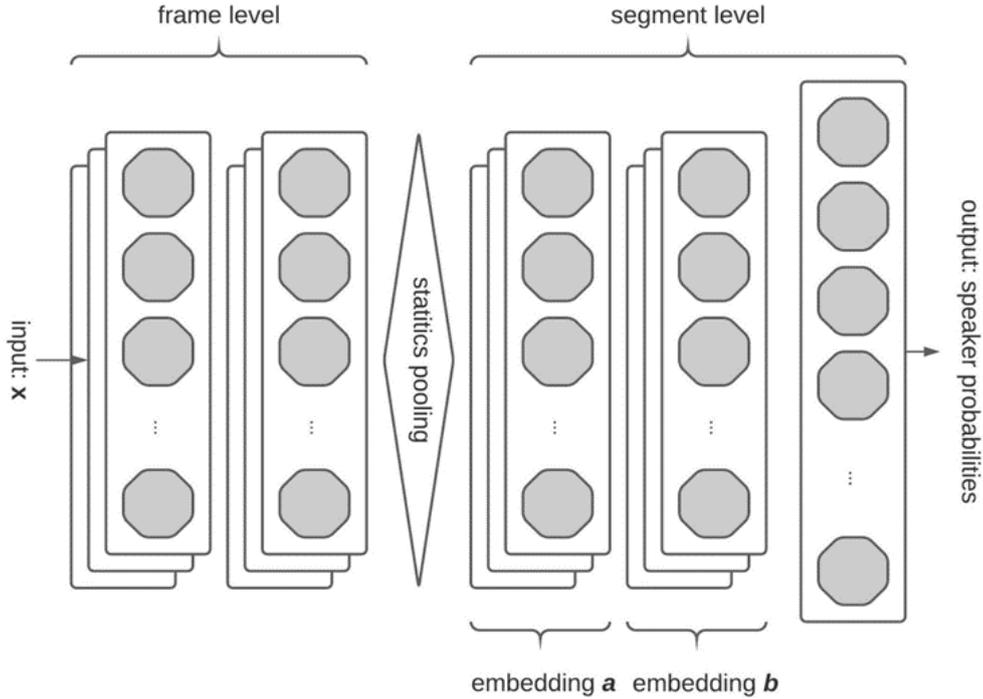

*Figure 1. The x-vector DNN embedding architecture in* (Snyder et al., 2017). *The two parts: frame level (with the 5 frame layers) and segment level (with segment6, segment7 and softmax).*

The first five layers operate on speech frames, with small temporal context centered at the current frame *t*. For example, the frame indexed as 3 sees a total of 15 frames, due to the temporal context of the earlier layers. After training with speaker ids as target vectors, the output of layer segment6 ('x-vector') is used as input to a classifier.

ECAPA-TDNN

The ECAPA-TDNN model is the extension of the x-vector model architecture in three ways (Desplanques et al., 2020): channel- and context-dependent statistics pooling, 1-Dimensional Squeeze-Excitation Res2Blocks (1D SE-Res2Block) and multi-layer feature aggregation and summation. The channel- and context-dependent statistics pooling enables the network to focus more on speaker characteristics that do not activate on identical or similar time instances, e.g. speaker-specific properties of vowels versus speaker-specific properties of consonants. Using the SE-Res2Block (taken from the field of computer vision), the limited frame context of the x-vector (15) is extended to global properties of the recording. The multi-layer feature aggregation means that not only the activation of the selected deep layer is used as a feature map (as in x-vector), but the shallower layers (here: SE-Res2Blocks) are also concatenated, because they also hold information about the speaker identity. The architecture is shown in Figure 2.

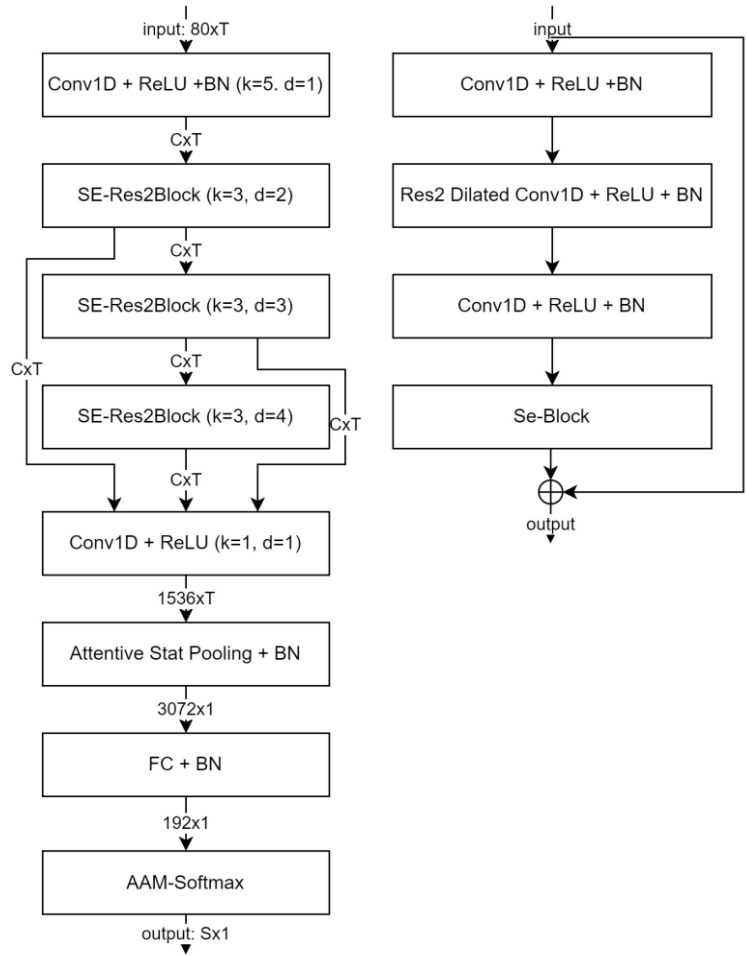

*Figure 2. The ECAPA-TDNN layer architecture and its SE-Res2Block.*

# Methods

## Embedding Models

### The x-vector

The dimension of the x-vectors was set to 512 and the input was 24 mel-frequency band energies. The training was done for 35 epochs with early stopping for which the criteria was the minimum loss that was measured on a validation set. The training was done with Adam optimizer with a starting learning rate of 0.001. The x-vector model pre-trained on the VoxCeleb dataset was downloaded from Huggingface and all custom-trained models followed the same recipe.

### ECAPA-TDNN

Following the recipe of the model pre-trained on the VoxCeleb dataset[3], the dimension of the extracted embedding vector in the case of custom-trained models was 192 and the input was 80 mel-frequency band energies. The training was done for 35 epochs with early stopping for which loss was measured on a validation set. The training was done with Adam optimizer with a starting learning rate of 0.001.

---

[3] https://huggingface.co/speechbrain/spkrec-ecapa-voxceleb

## Cosine Distance and Enrollment

In order to evaluate similarities of the extracted embedding vectors for sample pairs, cosine distance was used. The cosine distance is simply computing the normalized dot product of target and test vectors ($w_{target}$ and $w_{test}$), which provides a match score:

$$CDS(w_{target}, w_{test}) = \frac{w_{target} \cdot w_{test}}{||w_{target}|| \cdot ||w_{test}||}$$

If multiple samples are available for a known speaker, it is commonly advantageous to create an enrollment vector or model from them. Due to the usage of cosine distance in the present study, the embedding vectors were averaged per speaker to get a mean embedding vector for each speaker. In the results we compared the performance with and without using speaker enrollment. Speaker enrollment was done by averaging the embedded vectors per speaker on the session 1 samples (known speakers). The average vectors were then compared to all of the session 2 samples (unknown speakers).

## LR score calculation

To calculate LR scores, logistic regression was used (implemented in the Python sklearn package). Cosine distances calculated for sample pairs were arranged according to the same speaker and different speaker labels and logistic regression models are trained. The output of the logistic regression model is the probability of the same speaker decision. This enables the calculation of Eq. 1, where the probability of different speaker origin is 1-P(H$_{so}$). Figure 3 shows an example of a trained logistic regression model. Distribution of same and different origin vector pairs is shown in blue and yellow, respectively.

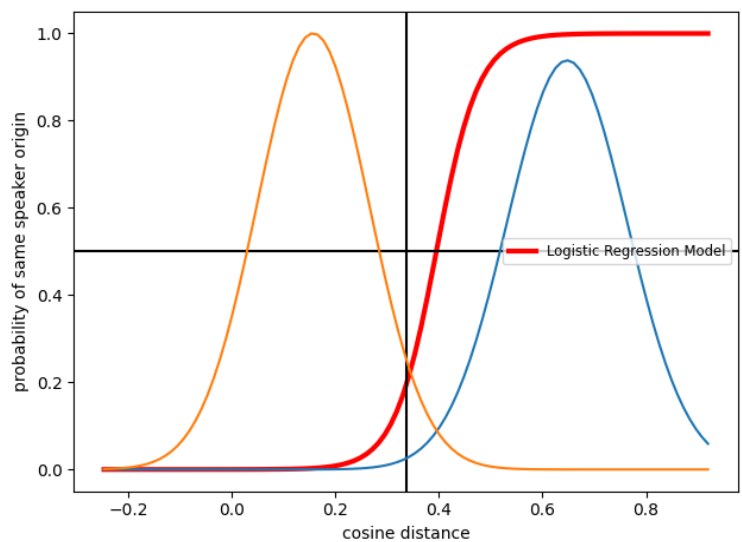

*Figure 3. A trained logistic regression model example. Blue and yellow lines show the distributions of cosine distances of embedding vector pairs of same and different speaker origin, respectively.*

# Datasets

The evaluation according to the forensic voice comparison performance metrics were done on two datasets created for forensic purposes (multiple speaking styles per speaker, multiple recording sessions per speaker): the Hungarian ForVoice120+ corpus and the Australian English AusEng (Morrison et al., 2015) dataset. They followed the protocol specified by Morrison and his colleagues (Morrison et al., 2012). Both datasets contain multiple speech tasks per speaker each modeling a different speaking style and (at least) two recording sessions per speaker with at least two weeks' delay. See Table 2 for dataset descriptions. The ForVoice120+ dataset contains 120 speakers, representing a dataset for a low-resource language. The AusEng corpus contains more than 500 speakers. Both corpora contain three speech tasks: free dialogue, information exchange and monologue (simulating interrogation). The datasets were split into multiple parts. 40 speakers of the ForVoice120+ were used for LR calibration and the remaining 80 speakers were used for evaluation. Because the ForVoice120+ contains only a limited number of speakers, the Hungarian speaker embedding x-vector and ECAPA-TDNN models were trained on different samples, not explicitly made for forensic purposes: BEA (Neuberger et al., 2014), MRBA (Vicsi et al., 2004) and newly recorded samples with read text and free speech. 632 speakers were used for the training altogether, the total duration of the speech was 27.31 hours. In the case of the AusEng dataset, 395 speakers were randomly selected for embedding model training, 80 speakers for LR calibration and 80 for evaluation.

*Table 2. Metadata for the ForVoice120+ and AusEng datasets*

| dataset | total number of recordings | number of recording sessions | number of speakers | (male/female) | total speech length | number of recordings per speaker |
|---|---|---|---|---|---|---|
| ForVoice120+ | 720 | 2 | 120 | 59 / 61 | ~32 hours | 6 |
| AusEng | 3899 | 2 or 3 | 555 | 239 / 316 | ~311 hours | 3-9 |

Beside embedding models trained on the Hungarian dataset and the AusEng corpus, pre-trained models on the VoxCeleb2 (Nagrani et al., 2020) corpus (available on the Huggingface repository) were also used for embedding feature vector extraction, because the corpus contains more than 7000 speakers and represents the largest model available for speaker recognition. The two English datasets are representing a language that is commonly available and has many resources: VoxCeleb dataset containing thousands of speakers. During the embedding model training, the used datasets were split into training, validation and test sets in 60-20-20% ratios. Early stopping criteria was measured on the validation set and the test set was used to check against overfitting (comparing the results on the test set and the validation set).

## Database splitting

The recordings of the dataset used for Hungarian embedding model training, ForVoice120+ and the AusEng datasets were split into multiple parts with various lengths. The possible duration of a part was {2,3,4,5,6,7,8,9,10} seconds. The number of samples were (almost) evenly distributed according to the durations. First, all silence parts were removed from the recordings, then splitting was done with 10% overlap between adjacent parts. There was no influence on the sample lengths of the VoxCeleb2 dataset, because pre-trained models were used in those cases. The final distributions of the sample durations are shown in Figure 4.

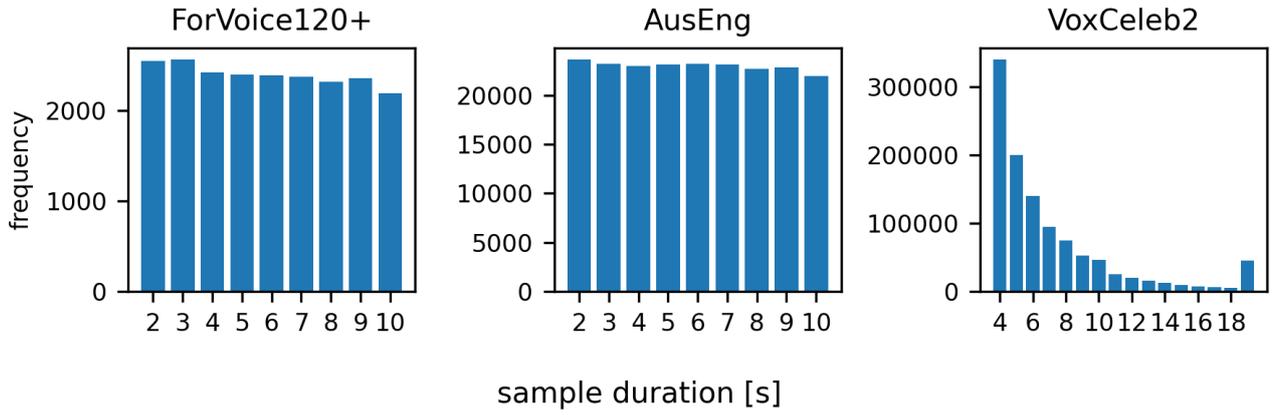

*Figure 4. Distributions of sample durations in the datasets used for the study*

Augmentation

Following the recipe of the pre-trained models available on Huggingface, data augmentation was applied during model training in the case of the Hungarian samples and the AusEng dataset. Samples were extended with time-distorted (duration was scaled with factors 0.95 and 1.05) and noise distorted (with 15 dB white noise) variants. According to the results reported on the pre-trained models, this augmentation increases the robustness of the models. In order to be able to compare the newly trained models with these pre-trained ones, the same augmentation was used.

Evaluation metrics

The outputs of the different model setups were evaluated by equal error rate (EER) of speaker verification (EER is the level where false acceptance rate and false rejection rate are equal, commonly used in biometric security systems) and log-likelihood-ratio cost (Cllr, Eq. 2) (Van Leeuwen and Brümmer, 2007), defined as

Eq. 2. $$Cllr = \frac{1}{2}\left(\frac{1}{N_{so}}\sum_{i=1}^{N_{so}}\left(1+\frac{1}{LR_{so_i}}\right) + \frac{1}{N_{do}}\sum_{j=1}^{N_{do}}\left(1+LR_{do_j}\right)\right)$$

where $N_{so}$ and $N_{do}$ are the number of same-origin and different-origin comparisons and $LR_{so}$ and $LR_{do}$ are the likelihood ratios derived from same-origin and different-origin comparisons. Cllr is a function measuring the balance of LR scores of same-origin and different-origin comparisons. Ideal same-origin and different-origin comparisons have logLR>0 and logLR<0, respectively. Incorrect (not as ideal as the mentioned inequalities) produce a higher Cllr. The better the performance of a forensic comparison system, the more correct LR values are produced, the lower Cllr is achieved, supplying the evidence magnitude. Beside Cllr, the minimum Cllr is also reported, which is the generalization of the original cost function and produces application independent Cllr values (Brümmer and Du Preez, 2006). Results are also shown on tippet plots, displaying the proportion of correctly identified same and different speaker origin pairs (commonly used visualization in forensic comparison).

Evaluation scenarios

Multiple phenomena were investigated:

- language mismatch,
- speaker enrollment,
- sample duration mismatch and
- speech task mismatch.

Datasets with different languages were applied to train embedding models, calibrate LR scores and evaluate speaker verification. This is done without and with enrollment. The best performing scenario was broken down into utterance duration and speech tasks to see their effect on forensic voice comparison. Embedding model training and LR score calibration were done on samples from all sessions. During the evaluation, samples were compared from different sessions: session 1 samples were used as known speaker samples and session 2 as unknown speaker samples. Sessions are recorded with at least two weeks' delay.

In the best performing dataset combination case, the results are broken down according to sample durations and speech tasks. The $Cllr_{min}$ and EER values (without enrollment) are organized into matrices of the examined phenomena. In the case of sample durations, rows and columns are the results calculated according to the durations from 2 to 10 seconds. A value of a cell was calculated by filtering the sample-pair comparisons according to the known and unknown speaker sample lengths. In the case of speech tasks, the rows and columns of the matrices are the task numbers (1: free dialogue, 2: information exchange, 3: monologue) and the cells contain the results of a given task pair. Using speaker enrollment, the single sample lengths and speech task were not applicable for the known speakers, because the speaker vectors were averaged through all session 1 samples. The results are therefore vectors in this case. However, this does not pose a problem, because this represents the real-life situation, when multiple recordings can be acquired from a suspect (known speaker) and the enrollment can be done.

# Results

## Effect of languages used for model training and LR calibration

The results using different dataset combinations without speaker enrollment is shown in Table 3. ECAPA-TDNN models outperformed the x-vectors in all cases. Significant decrease can be found in all metrics. The best performing combinations are when the VoxCeleb dataset was used for embedding model training (pre-trained models): 3.1% EER and 0,122 $Cllr_{min}$ values for ECAPA-TDNN. Compared to the state of art results on short utterance comparisons, this is a good value. The LR calibration set did not cause any difference. Comparing the evaluation sets, the language difference also did not decrease the performance. Even a little bit higher metric values were found using the same language (although different dialect). The tippet plot of the best performing case is shown in Figure 5.

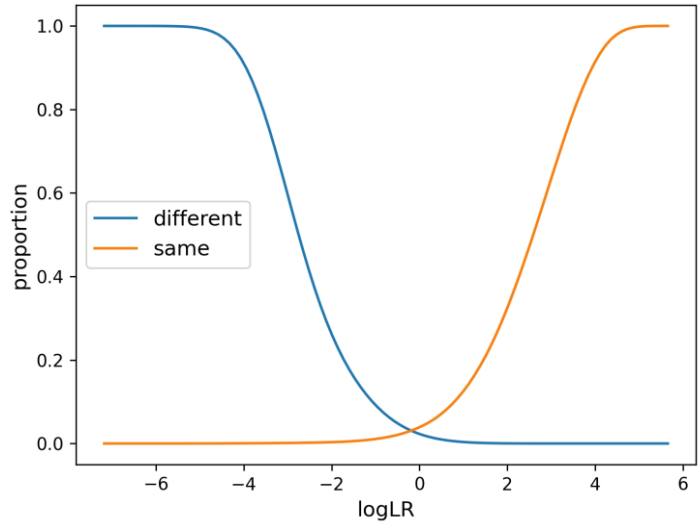

*Figure 5. Tippet plot of dataset combination (ECAPA-TDNN pretrained on VoxCeleb and LR calibration on ForVoice120+) with lowest Cllr$_{min}$.*

*Table 3. Speaker verification results obtained with models of different dataset combinations*

| Ealuation language/dataset | Embedding model language/dataset | LR calibration language / dataset | Model | Cllr | Cllr$_{min}$ | Cllr$_{cal}$ | EER |
|---|---|---|---|---|---|---|---|
| Hungarian / ForVoice120 | Hungarian | Hungarian / ForVoice120 | x-vector | 0,632 | 0,601 | 0,031 | 0,189 |
| | | | ECAPA-TDNN | 0,405 | 0,401 | 0,005 | 0,116 |
| | English / AusEng | Hungarian / ForVoice120 | x-vector | 0,567 | 0,537 | 0,031 | 0,167 |
| | | | ECAPA-TDNN | 0,253 | 0,249 | 0,004 | 0,069 |
| | English / AusEng | English / AusEng | x-vector | 0,581 | 0,537 | 0,044 | 0,167 |
| | | | ECAPA-TDNN | 0,629 | 0,249 | 0,380 | 0,069 |
| | **English / VoxCeleb** | Hungarian / ForVoice120 | x-vector | 0,365 | 0,349 | 0,016 | 0,102 |
| | | | **ECAPA-TDNN** | **0,127** | **0,122** | **0,005** | **0,031** |
| | **English / VoxCeleb** | English / AusEng | x-vector | 0,381 | 0,349 | 0,032 | 0,102 |
| | | | **ECAPA-TDNN** | **0,168** | **0,122** | **0,046** | **0,031** |
| English / AusEng | English / AusEng | English / AusEng | x-vector | 0,616 | 0,575 | 0,041 | 0,183 |
| | | | ECAPA-TDNN | 0,184 | 0,182 | 0,001 | 0,048 |
| | English / VoxCeleb | English / AusEng | x-vector | 0,511 | 0,481 | 0,029 | 0,150 |
| | | | ECAPA-TDNN | 0,220 | 0,206 | 0,014 | 0,053 |

## Effect of enrollment

The same dataset combinations were repeated applying speaker enrollment. The result is shown in Table 4. The main tendencies (embedding vector technique differences, dataset used for embedding models, evaluation datasets) are the same as before. The ECAPA-TDNN outperformed the x-vector in this case also. Language differences did not cause performance degradation. Pre-trained models available on the VoxCeleb dataset are the best performing ones here too. The lowest EER is 1% with 0,045 Cllr$_{min}$. The tippet plot of this case is shown in Figure 6.

*Table 4. Speaker verification results obtained with models of different dataset combinations by speaker enrollment*

| Evaluation language/dataset | Embedding model language/dataset | LR calibration language / dataset | Model | Cllr | Cllr$_{min}$ | Cllr$_{cal}$ | EER |
|---|---|---|---|---|---|---|---|
| **Hungarian / ForVoice120** | Hungarian | Hungarian / ForVoice120 | x-vector | 0,517 | 0,411 | 0,105 | 0,123 |
| | | | ECAPA-TDNN | 0,190 | 0,183 | 0,006 | 0,049 |
| | English / AusEng | Hungarian / ForVoice120 | x-vector | 0,609 | 0,343 | 0,265 | 0,104 |
| | | | ECAPA-TDNN | 0,115 | 0,110 | 0,005 | 0,029 |
| | English / AusEng | English / AusEng | x-vector | 0,458 | 0,343 | 0,115 | 0,104 |
| | | | ECAPA-TDNN | 0,378 | 0,110 | 0,267 | 0,029 |
| | **English / VoxCeleb** | **Hungarian / ForVoice120** | x-vector | 0,248 | 0,191 | 0,058 | 0,053 |
| | | | **ECAPA-TDNN** | **0,050** | **0,045** | **0,006** | **0,010** |
| | **English / VoxCeleb** | **English / AusEng** | x-vector | 0,222 | 0,191 | 0,031 | 0,053 |
| | | | **ECAPA-TDNN** | **0,067** | **0,045** | **0,022** | **0,010** |
| English / AusEng | English / AusEng | English / AusEng | x-vector | 0,472 | 0,358 | 0,114 | 0,104 |
| | | | ECAPA-TDNN | 0,064 | 0,061 | 0,004 | 0,016 |
| | English / VoxCeleb | English / AusEng | x-vector | 0,324 | 0,289 | 0,035 | 0,085 |
| | | | ECAPA-TDNN | 0,093 | 0,084 | 0,009 | 0,020 |

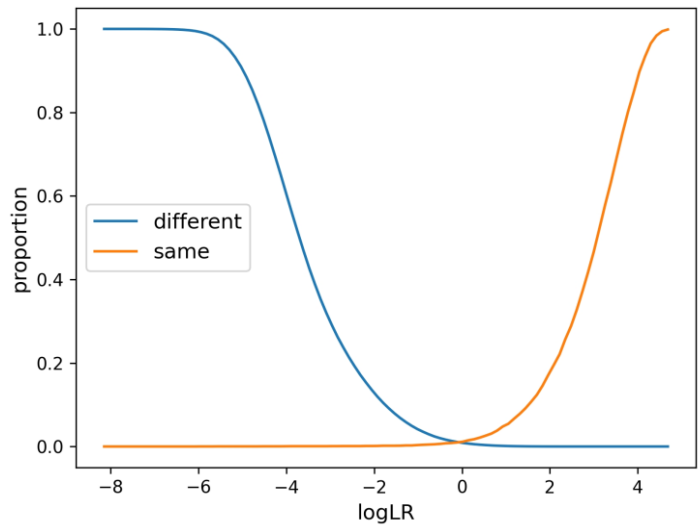

*Figure 6. Tippet plot of dataset combination (ECAPA-TDNN pretrained on VoxCeleb and LR calibration on ForVoice120+) with lowest Cllr$_{min}$ for enrollment.*

## Effect of sample duration

From the two best performing cases, the Hungarian was selected to examine sample duration effects due to its low-resource nature. Figures 7 and 8 show the Cllr$_{min}$ and EER values (without and with enrollment). As was suspected, the longer the duration of the sample the better the results are. When enrollment is not applied, instead of the global 3.1% EER, the 10vs10 case reaches 1.1% EER. On the other hand, the shortest case (2vs2) is 5%. However, considering that this means comparing samples of 2 seconds of the unknown and the known speakers, 5% can be an acceptable result in real life. By applying speaker enrollment, samples with 10 second duration reach 0.2% EER. The shortest samples go up to 1.6%.

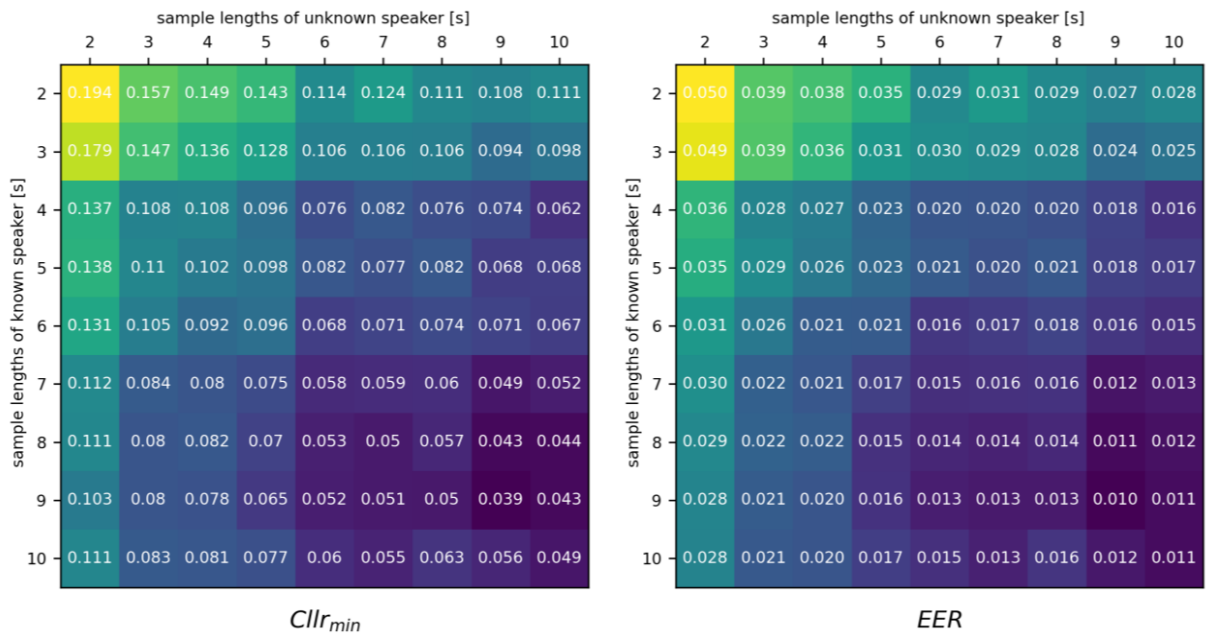

*Figure 7. Heatmap of $Cllr_{min}$ and EER values depending on sample duration without speaker enrollment. ECAPA-TDNN models trained with VoxCeleb, LR calibration done on ForVoice120+.*

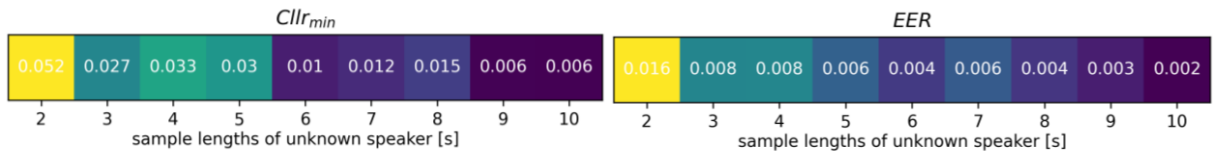

*Figure 8. Heatmap of $Cllr_{min}$ and EER values depending on sample duration with speaker enrollment. ECAPA-TDNN models trained with VoxCeleb, LR calibration done on ForVoice120+.*

## Effect of speaking tasks

The results of the best performing dataset combination case were also broken down into speech task combinations. The $Cllr_{min}$ and EER values are shown in Figures 9 and 10. Based on the results, task 3 vs task 3 (monologue, describing the events of the previous day of the speaker, EER: 1.8%) has the lowest EERs and the highest values are achieved in cross-task setups (task 1 vs task 2, EER: 3.6%). In the case of speaker enrollment, the results show the same tendency. Using the monologue gives the best results (0.8% EER) and the information exchange gives the worst, although just slightly higher (1.2% EER).

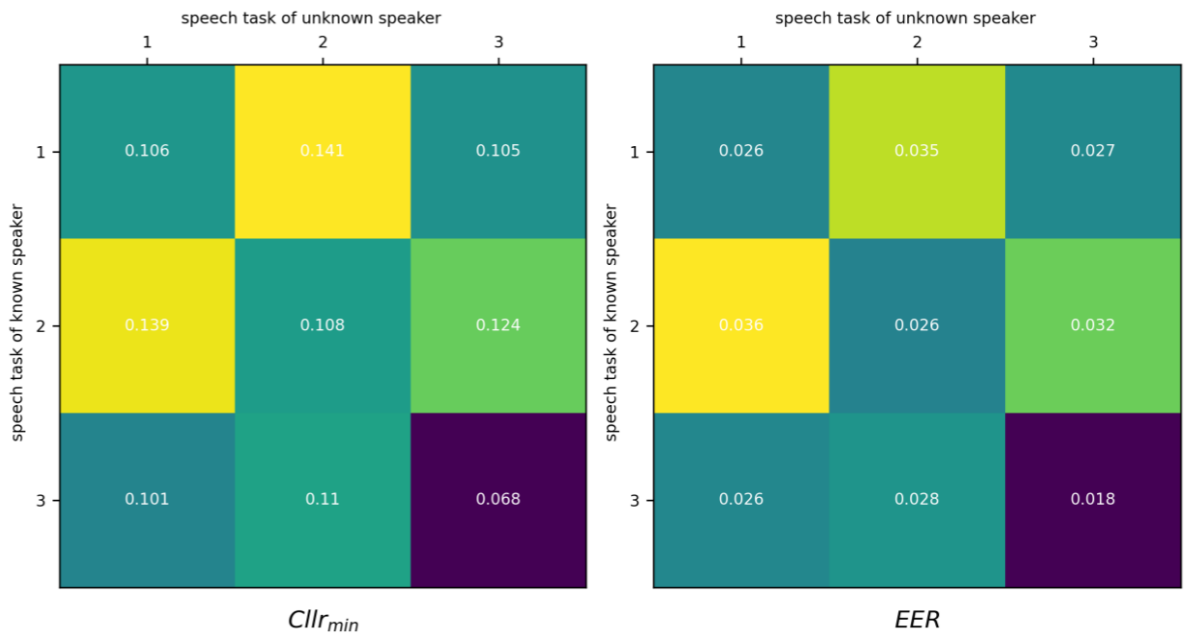

*Figure 9. Heatmap of $Cllr_{min}$ and EER values depending on speech task without speaker enrollment. The ECAPA-TDNN models were trained on VoxCeldeb, LR calibration was done on ForVoice120+.*

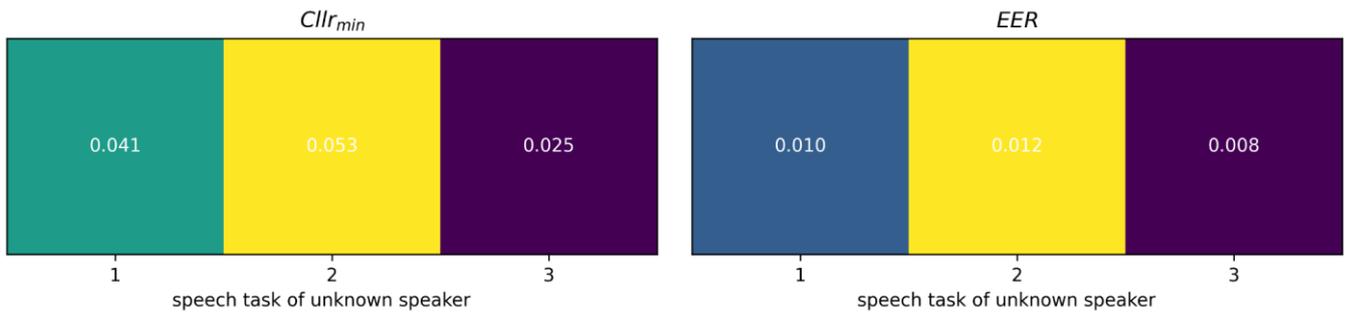

*Figure 10. Heatmap of $Cllr_{min}$ and EER values depending on speech task with speaker enrollment. The ECAPA-TDNN models were trained on VoxCeleb, LR calibration was done on ForVoice120+.*

# Discussion and Conclusion

In the present study, language mismatch effects were examined from a forensic voice comparison perspective using speaker embeddings. The goal here was to evaluate whether language differences matter in voice comparison and to investigate if a model pre-trained on a language different from the target samples can be used for forensic voice comparison. The results show that this is so. The lowest EER (3.1% and 1.0% without and with speaker enrollment, respectively) was obtained with the model pre-trained on the VoxCeleb dataset and evaluating it on the Hungarian ForVoice120+ corpus. This was even better than evaluating the model on the English AusEng dataset. Therefore, it can be stated that language difference does not deteriorate the performance of the given technique based on deep learning embeddings. The language of the corpus used for LR calibration does not seem to affect the performance. From the two applied deep learning architectures, models structured by the ECAPA-TDNN architecture perform better in all corpus combinations than the x-vector. This is consistent with what Desplanques and colleagues reported (Desplanques et al., 2020).

Considering the speaker enrollments, additional performance increase can be achieved. The results show that (unsurprisingly) if multiple samples are available from the suspect, it is better to apply speaker enrollment (in this study, the average of the embedding vectors) to compare the offender's voice sample. In the case of the best performing model 2.1% absolute EER can be achieved (from 3.1% to 1.0%). Of course, this is only available if there are multiple recordings from the suspect, but during an investigation process, recordings can be acquired on purpose.

Breaking down the results according to sample lengths and speaking styles (simulated by different speech tasks), more details can be revealed. Sample length analysis shows that the longer the duration of the sample in question the better the performance is, as one would naturally expect. Comparing sample pairs with 2 seconds durations, 5% EER can be achieved at best, while using 10 seconds-long samples, this goes down to 1.1%. With enrollment (comparing a single sample of the offender to the average vector of the suspect) this can be further improved: 0.5% EER in the case of 10 seconds. This implies that models pre-trained on samples with a different language can be used in practice, but the longer the sample, the better the performance. Also, it is recommended to record multiple samples from the suspect and use an averaged embedding vector.

Results according to the speech task show that there is a slight gain if the same speaking style is used in the compared samples (at least spontaneity would be the same) if enrollment is not possible. However, if multiple recordings are present from the suspect, there is no real difference if various speaking styles are applied, and it doesn't really matter.

The results in the study show that the automatic sample-wise forensic voice comparison technique applied in the present study can be used in practical, real-world scenarios. This is useful if only a single sample is available from the offender. (There can be multiple samples available from the suspect, as speaker enrollment is also investigated here.) The achieved Cllr and EER values cannot be compared directly to other studies (due to the differences in the applied datasets). The goal here was not to evaluate the speaker embedding models (they were already evaluated in other studies), but to examine the language mismatch properties of a possible voice comparison system.

The dataset used for evaluation is in a low-resource language (Hungarian). These types of languages often do not have enough speech samples to train a deep learning model and to calibrate a LR framework. The results obtained here show that models pre-trained on a different language can indeed be used on the target language (which is also linguistically distant). These pre-trained models (even available online) may be applied to speaker verification universally independent of the language of the samples in question. This may be very useful if this language is a low-resource language, where few samples are available for forensic purposes (neither proper forensic evaluation is possible, nor a deep learning model can be trained).

# Acknowledgement

The work was funded by project no. FK128615, which has been implemented with the support provided from the National Research, Development and Innovation Fund of Hungary, financed under the FK_18 funding scheme.